**Effectiveness of Self-Assessment Software to Evaluate Preclinical Operative Procedures**

Qi Dai, B.S.[1]; Ryan Davis, D.D.S, M.S.[2]; Houlin Hong, M.S.[1]; Ying Gu, D.D.S., Ph.D.[1]

[1]Department of General Dentistry, School of Dental Medicine, Stony Brook University; Stony Brook, NY, USA;

[2]Department of Community Health & Social Sciences, Graduate School of Public Health & Health Policy, City University of New York, New York City, NY, USA..

Direct correspondence: Ying Gu, Department of General Dentistry, School of Dental Medicine, Stony Brook University, Stony Brook, NY 11794, Tel: 631-632-8723, Email: ying.gu@stonybrookmedicine.edu.

**Effectiveness of Self-Assessment Software to Evaluate Preclinical Operative Procedures**

**Abstract:**

**Objectives:** To assess the effectiveness of digital scanning techniques for self-assessment and of preparations and restorations in preclinical dental education when compared to traditional faculty grading.

**Methods:** Forty-four separate Class I (#30-O), Class II (#30-MO) preparations, and class II amalgam restorations (#31-MO) were generated respectively under preclinical assessment setting. Calibrated faculty evaluated the preparations and restorations using a standard rubric from preclinical operative class. The same teeth were scanned using Planmeca PlanScan intraoral scanner and graded using the Romexis® E4D Compare Software. Each tooth was compared against a corresponding gold standard tooth with tolerance intervals ranging from 100µm to 500µm. These scores were compared to traditional faculty grades using a linear mixed model to estimate the mean differences at 95% confidence interval for each tolerance level.

**Results:** The average Compare Software grade of Class I preparation at 300µm tolerance had the smallest mean difference of 1.64 points on a 100 points scale compared to the average faculty grade. Class II preparation at 400µm tolerance had the smallest mean difference of 0.41 points. Finally, Class II Restoration at 300µm tolerance had the smallest mean difference at 0.20 points.

**Conclusion:** In this study, tolerance levels that best correlated the Compare Software grades with the faculty grades were determined for three operative procedures: class I preparation, class II preparation and class II restoration. This Compare Software can be used as a useful adjunct method for more objective grading. It also can be used by students as a great self-assessment tool.

**Introduction:**

Students' first experiences practicing on teeth is typically seen in the operative dentistry I course where they are expected to prepare numerous teeth and then restore them to normal function. This is these students' first experiences practicing with handpieces and the most crucial in developing proper manual dexterity skills and techniques for their future careers in dentistry. Often students are unsure of the quality of their preparations and ultimately rely on the faculty or other students for feedback.

For generations, the students relied on faculty for both feedback on their performance and for grading competency exams. While faculty evaluations are critical for improvements, they are not always available for timely feedback due to them with other academic responsibilities or if the student is practicing outside of designated class hours.[1,2] Limitations to faculty feedback can lead to several undesirable consequences including student frustration, confusion, and compromised education.[3] Thus, there is a need to identify alternative grading and evaluation methods to allow for self-evaluation to effectively improve the students' skills and understanding. Studies in the past have shown that certain technologies would be able to assist in evaluation of such procedures by eliminating a factor of subjectivity. Recent advances in dental technologies have made intraoral scanners increasingly more powerful and common to be seen throughout the field.[4] The software alongside these scanners have also shown great improvements allowing students to go as far as evaluate preparations for undercuts present in their preparations.[5]

The Planmeca Romexis® E4D Compare Software (Compare Software) is designed as a teaching aid, allowing students to take two scanned models using intraoral dental scanners and overlay them in real time to visualize their differences. The software can identify key features shared by the two models and auto-align them with precision. This software can then display differences between the two models, color coding how the two are different, even being able to provide a percentage match between the two models. Two dimensional slices are available to see exactly where the two models diverge as seen in *fig. 1*. The most important feature is the ability to define an area of the models to be compared, allowing for precise measurements at the area of interest.

The Planmeca Romexis® E4D Compare Software was previously used to digitally evaluate wax-ups and crown preparations which gave feedback from dental students.[4,7-9] This software can provide unbiased feedback and allowing for a detailed view of where improvements can be made. To fully utilize this technology, there is a need to optimize the setup of the software. Depending on the tolerance value set, variable from 1μm to 500μm, the differences between the two models will be either magnified or hidden, reducing its ability to critically assess the preparations. The goal of this study was to identify the ideal tolerance value for grading class I preparations, class II preparations and class II restorations in typodont teeth. This study was done by comparing the Compare Software grades generated for each of these procedures compared to faculty assessments of the same preparation and restoration.

**Materials and Methods:**

This study was evaluated by Stony Brook University Institutional Review Board by either the Common Rule or FDA regulations, and was deemed to exempt determined to be eligible for exemption .

Forty-four of each class I preparations, class II preparations and class II restorations (132 teeth total) were generated by first year dental students under competency conditions as part of their Operative Dentistry I course at Stony Brook University School of Dental Medicine. Plastic typodont teeth (Ivorine®, Columbia Dentoform, PA) were utilized for this course and the students had full access to their operative bur blocks containing the carbide burs - 330, 245, 556, and any others which they desired. Students were instructed to prepare ideal preparations and restorations. The teeth used for each operative procedure were tooth number 30 (Mandibular right first molar) occlusal for the class I preparation (#30-O), number 30 mesio-occlusal (#30-MO) for the class II preparation and number 31 (Mandibular right second molar) MO for the class II amalgam restoration (#31-MO).

These teeth were graded by five calibrated faculty members using a standard rubric for operative dentistry course on a scale from 0-100 in five-point intervals. During the calibration process, each faculty was presented with standard preparations or restorations from previous years and presented with proper grades for each tooth. Once the faculty were familiarized with the standard of the grading scheme, they were considered calibrated. The grading was done double-blinded to eliminate any subjective factors. These teeth were graded against an ideal/standard preparation or restoration. The following aspects were considered when grading a preparation: over-extension, under-extension, convenience form, definition of the walls and floors, level of depth, roughness, and smoothness. The following aspects were considered when grading a restoration: form and anatomy of the cusps, grooves, ridges, and fossae, contours, surface smoothness, proximal contacts, and overhang. Following faculty grading, all the teeth were collected for later analysis using the Romexis® E4D Compare Software.

Using the same typodont for all teeth, the preparations and restorations were digitally scanned using the Planmeca PlanScan intraoral scanner and evaluated by the Romexis® E4D Compare Software. For the class I preparations, all adjacent teeth were present; for the class II preparations and restorations, the mesial tooth was removed to allow for visual access to the proximal box during scanning. Teeth were placed one at a time into the same typodont with the same adjacent teeth to eliminate any inconsistencies between scans and only the tooth being graded was changed. The STL file for each tooth was exported and saved for later digital grading.

A single ideal/standard preparation for each of the two preparations was then scanned into the program using the same experimental setup as the graded teeth. For the class II restorations, an untouched typodont tooth was used as the gold standard to compare to the graded restorations. Each of the digital models was trimmed to contain the same teeth and section of the typodont.

Using the Compare Software, the respective ideal preparation or restoration and the corresponding graded tooth were digitally aligned using the software's alignment feature which matches common anatomical features of the adjacent teeth. Once proper alignment of the two models was confirmed, the outline of the preparation was defined on the gold standard preparation to ensure consistent grading areas between teeth (*fig. 2*). The Compare Software allows for variable tolerance intervals for similarities between the ideal and experimental teeth on a scale from 0μm to 500μm. If the two models are further apart than the tolerance defined, it will be marked as not matching and if it is within the tolerance, it will be marked as matching. The displayed models will show any discrepancy between the two models by a particular color, green displays a match between the two models within the defined tolerance, red displays over-reduction and blue displays under-reduction. Each of the generated compare reports were then measured for percent match at the tolerances of 0μm, 100μm, 200μm, 250μm, 300μm, 350μm, 400μm and 500μm. This procedure was performed for all the teeth in three procedure groups.

For this study, the numerical value for the percentage match between the two models was recorded for each tolerance value. Summary statistics were first calculated including mean and standard deviation for both traditional and digital gradings. Linear mixed method was then used to the calibrated faculty grading to the digital gradings with an estimate of the mean differences between the two grading methods and 95% confidence intervals. The ideal tolerance value for grading each operative procedure was then selected. All statistical analysis was performed using SAS 9.4 (Cary, NC)

**Results:**

The mean scores with the traditional grading method of class I preparations, class II preparations and class II restorations were 80.45 (SD=5.79), 80.45 (SD=6.36), and 80.80 (SD=5.28), respectively. The descriptive statistics for digital grading with corresponding tolerance levels were summarized in *Table 1*, with the mean scores increasing as the tolerance level increases for all three groups.

The difference between the traditional grading and digital grading was significantly higher at the lowest tolerance of 100μm for all three operative procedures. At low tolerance values, it was found that the average digital grades given were lower than the faculty grading; while at the highest tolerance values, the average grades given from the Compare Software were higher than faculty grading. The mean differences between the faculty and digital gradings for the generated preparations and restorations are shown below (*fig. 3-5*).

The smallest mean difference between the traditional and digital grading methods was found at 300μm (Mean difference = 1.64, 95%CI= -1.33-4.60, p=0.27) for the class I preparations, 400μm (Mean difference = 0.41, 95%CI =-2.47-3.29, p = 0.78 ) for the class II preparations and 300μm (Mean difference = 0.20, 95%CI =-3.98-4.39, p = 0.92) for the class II restorations.

**Discussion:**

In this study, the lowest average grade generated by the Compare Software for all three evaluations was at the lowest tolerance level (100µm) as any minor deviation from the gold standard scan would be deemed as less "ideal" and hence a deduction in grades. On the contrary, the highest tolerance (500µm) gave too much leeway for errors in a subject and was less critical in the evaluation and not useful when students want to self-evaluate their skills. The specific tolerance level determined in this study for each preparation or restoration that is the closest to the faculty grading is an exceptional replacement for faculty feedback when they are not accessible. Students reported the feedback from computer-based programs as a useful tool during self-evaluation, especially when the faculty is not available.[11,12] Students even reported to perform better when using the Compare Software during practice, where they were provided access to digital scanning and evaluation on a volunteering basis[9]. However, the results were not significantly different from students who did not use this method due to the small sample size.

Our study found that the computer-based Compare Software can generate a grade/evaluation that is consistent and efficient for the preclinical operative procedures. It also eliminated any human bias or errors that may be involved during faculty grading, or the need of having multiple faculty to grade the same preparation to eradicate variations among graders.[10] The develop an accurate digital grading system, free of bias, is critical, it not only removes any subjective factors during the faculty grading, but also reduces the burden on existing faculty members. The results from our study provided a preliminary baseline for future development of an objective and accurate digital grading system.

Commission on Dental Accreditation (CODA) illustrated the importance of a dental graduates' ability to self-assess in its updated guidelines for dental accreditation in 2020,[13] It indeed is one of the most crucial trait a clinician should grasp from the 4-year dental education. Students cannot improve if they failed to reflect and assess themselves critically. Low performers often times overestimate their performance, whereas high performers tend to underestimate their performance, when compared to digital evaluation which thought to be accurate and unbiased.[14,15] Computer based scanning and evaluation is a great tool for a dental student to learn to self-assess and the cross-sectional visualization of the program is priceless when the student wants to look closely at the preparation and reflect on how to improve.[16-18] The ability of self-assessing will carry over from preclinical to clinical settings. Therefore, it is fundamental for dental students to develop this skill in a simulation setting.[19] In addition, it is found that students in clinical settings often underestimate their performance. Surprisingly, one study found that there is a weak correlation between preclinical and clinical performance judging the students on the same crown preparation on either typodont teeth or in patients', which suggested unsuccessful transfer of skills.[20] It is thought to be due to the inadequate practicing after completion of preclinical trainings which led to unpredictable results in clinical settings. Furthermore, in the simulation setting, the mannequin cannot completely duplicate the complications of the same procedures on a real patient, such as limited opening, tongue, gag-reflex, and saliva etc. Therefore, it is of uttermost importance for a student to become self-motivated to strive for the highest standard as there will be no supervision when practicing dentistry in the real world.

    The ideal interval settings found in this study would be a great starting point for anyone who wants to improve their preclinical skills but did not know how to interpret or criticize their performance based solely on a grading rubric. The Compare Software not only gives an objective evaluation of the student's performance, but also acts as a zoom-in image to show the student where the difference is in their performance compared to the standard. A grading rubric is a great tool to standardize grading process and attempt to eliminate any subjective influence.[21] However, it will never encompass the amount of information obtained from a digital scan, where the exact area of deficiency can be shown as an enlarged image on a computer screen to help students visualize their preparations or restorations directly. When it comes to practicing for operative dentistry, direct visualization is always easier to understand than reading the grading rubric on a piece of paper and trying to criticize one self's performance, especially in the early career of a dental student. In this study, placing a digital scanning of the student preparation and a gold standard that the student is trying to replicate side by side, or overlapping them, tells the student exactly where and how much it is different from the ideal preparation. Hence the student can improve at a faster pace than waiting for someone else to point out the inadequacy of the preparation or restoration. The benefit with this digital technology is apparent, though one should still refer back to the specific rubric established by each school since the standard does vary from school to school.

Students will be able to develop self-assessing skills with seniority, as it is a truly learned-process and takes practices to do so.[22] With the aid of digital comparison, it will take less time for students to learn to evaluate their own performance and grow faster as a clinician. Once the students improved their preclinical sensorimotor skills with practicing, they will have the option to narrow down the tolerance level to advance themselves with a more challenging standard. With the increasing number of dental school faculty vacancies, this would significantly improve students' learning experiences and relief the burden on existing faculty members.[23] Thus the faculty can better distribute their time in class for more advanced feedback that the computer would not be able to provide and use the digital software for more generic assessment.

On the other hand, in order to obtain accurate digital information for comparison, it is critical for one to be equipped with sufficient scanning skills. Because imprecise scanning does not give the system all the information regarding the student's preparation or restoration, thus creating erroneous grades and becoming misleading. Thankfully, at least 76% of all US dental schools started incorporating CAD/CAM technology into their pre-doctoral curriculum, according to a survey in 2015.[24] Consequently, it should not be compromising on student's busy schedule as they are required to learn this skill, nonetheless. Therefore, inaccuracies should be less of an issue in this matter.

Note that even though the Compare Software could generate a score that's comparable to faculty grading, it is still important to obtain feedback from faculty directly in terms of areas of improvements and weaknesses on one's preparation. There is no way to substitute faculty members entirely in this process, it is rather an aide in preclinical dental education.[25,26] This tool has to be used critically and carefully during self-assessing, and faculty feedbacks are still recommended.

**Conclusion:**

This study demonstrated that there is a high correlation between computer generated grades using the Romexis® E4D Compare Software and faculty grades of the preclinical preparations of #30 occlusal (Class I) and #30 mesio-occlusal (Class II), and amalgam restorations of #31 mesio-occlusal, with the tolerance settings at 300μm, 400μm, and 300μm, respectively. It is a great adjunct to preclinical education, as well as self-assessing. It greatly reduced the amount of time for students to obtain individualized feedback from faculty and thus reduce faculty workload in preclinical courses. It can be used confidently without any bias and is accessible to students at any time.

Accessed November 6, 2022.

Figures and Tables:

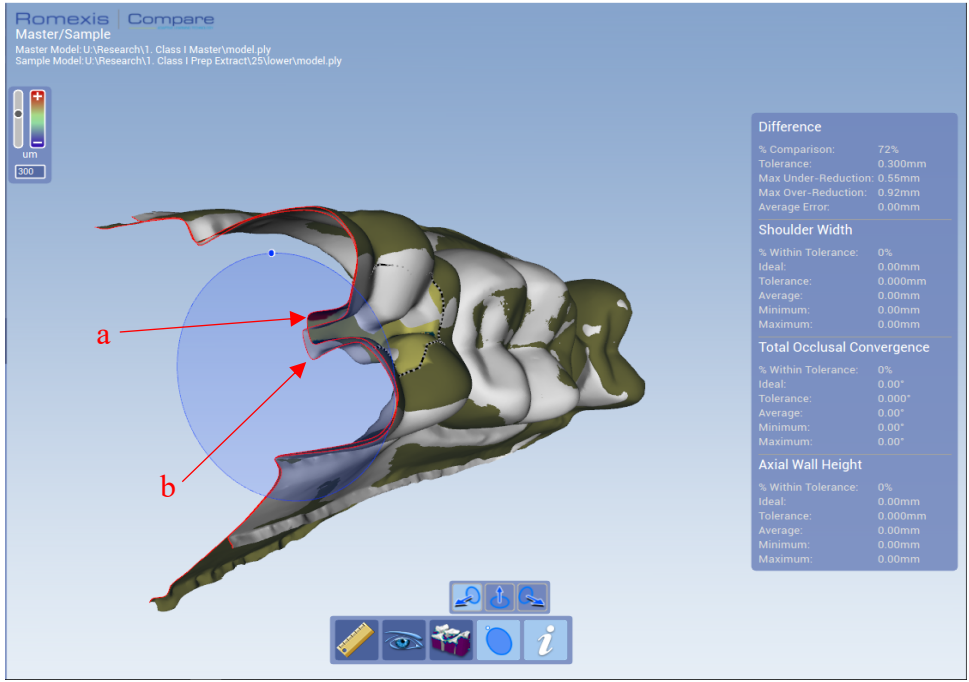

*Fig. 1.* Screenshot of E4D Compare software generated cross-section of a Class I preparation. a. the outline of prepared tooth being evaluated, b. the outline of the ideal preparation.

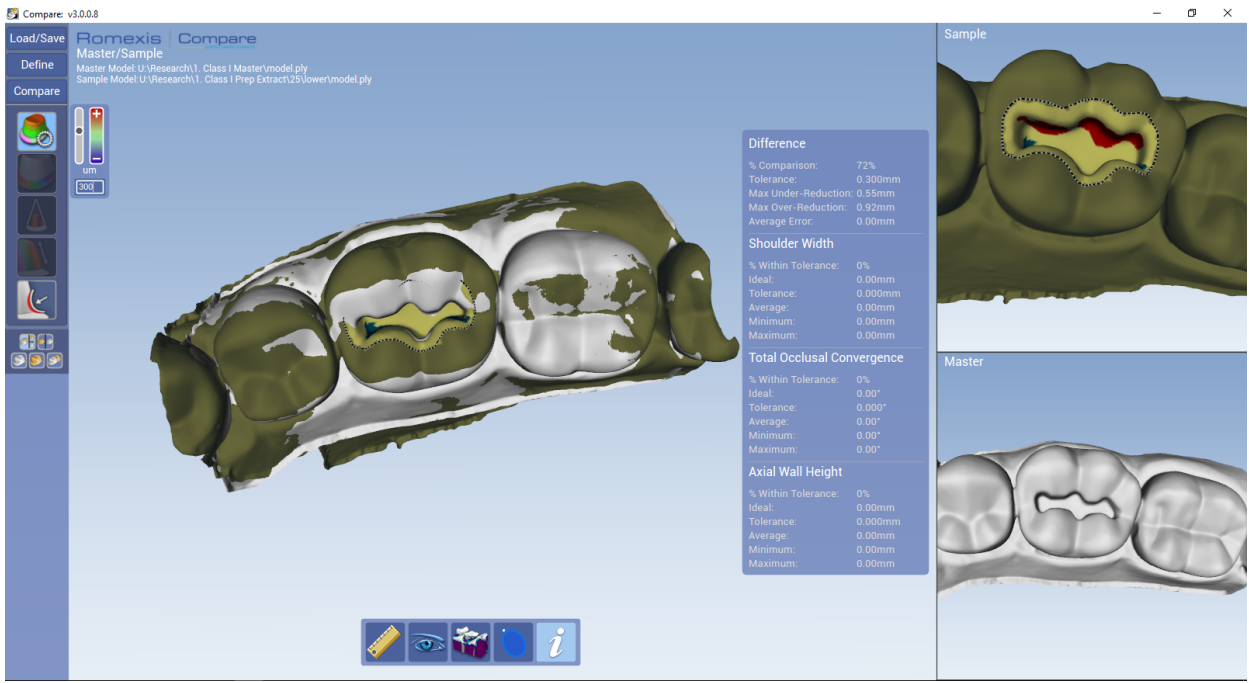

*Fig. 2.* Screenshot of E4D Compare software generated overlay of a Class I preparation with the master cast being the standard/ideal preparation. Red area means over-reduced and blue area means under-reduced when compared to the standard preparation.

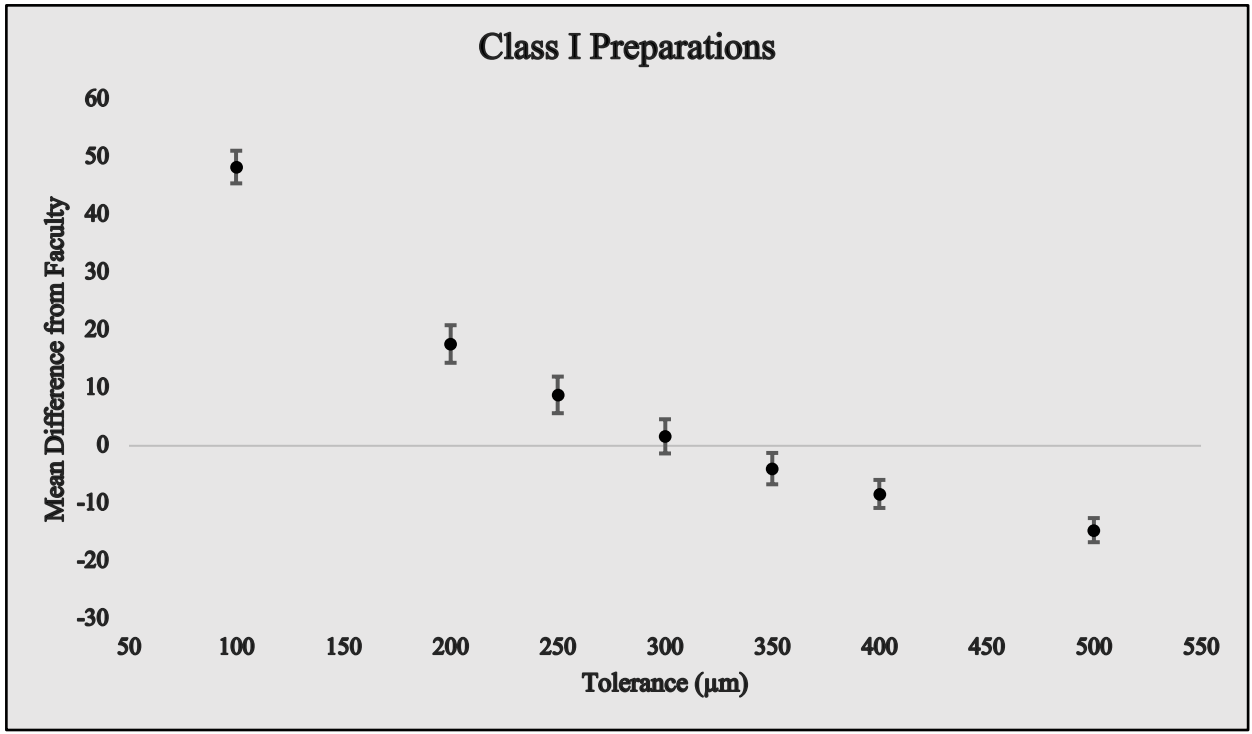

*Fig. 3*. The mean difference between faculty and machine grades for the Class I preparations at different tolerance levels.

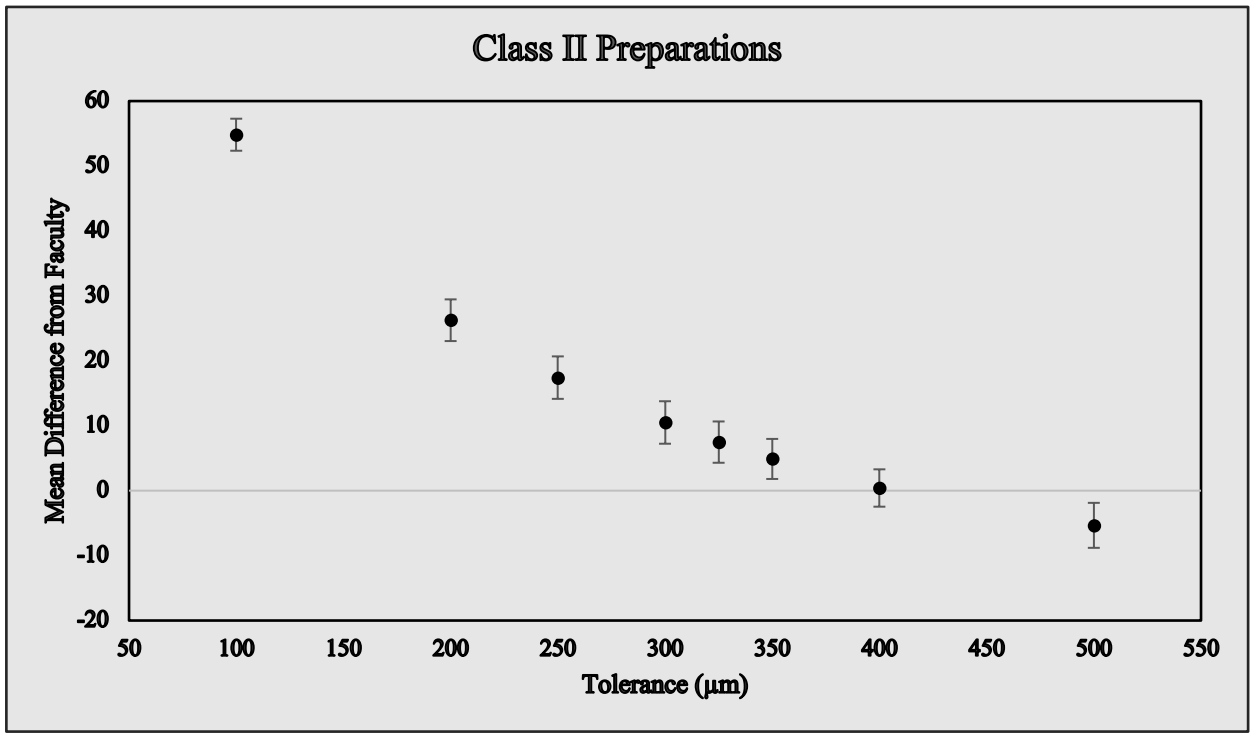

*Fig. 4*. The mean difference between faculty and machine grades for the Class II preparations at different tolerance levels.

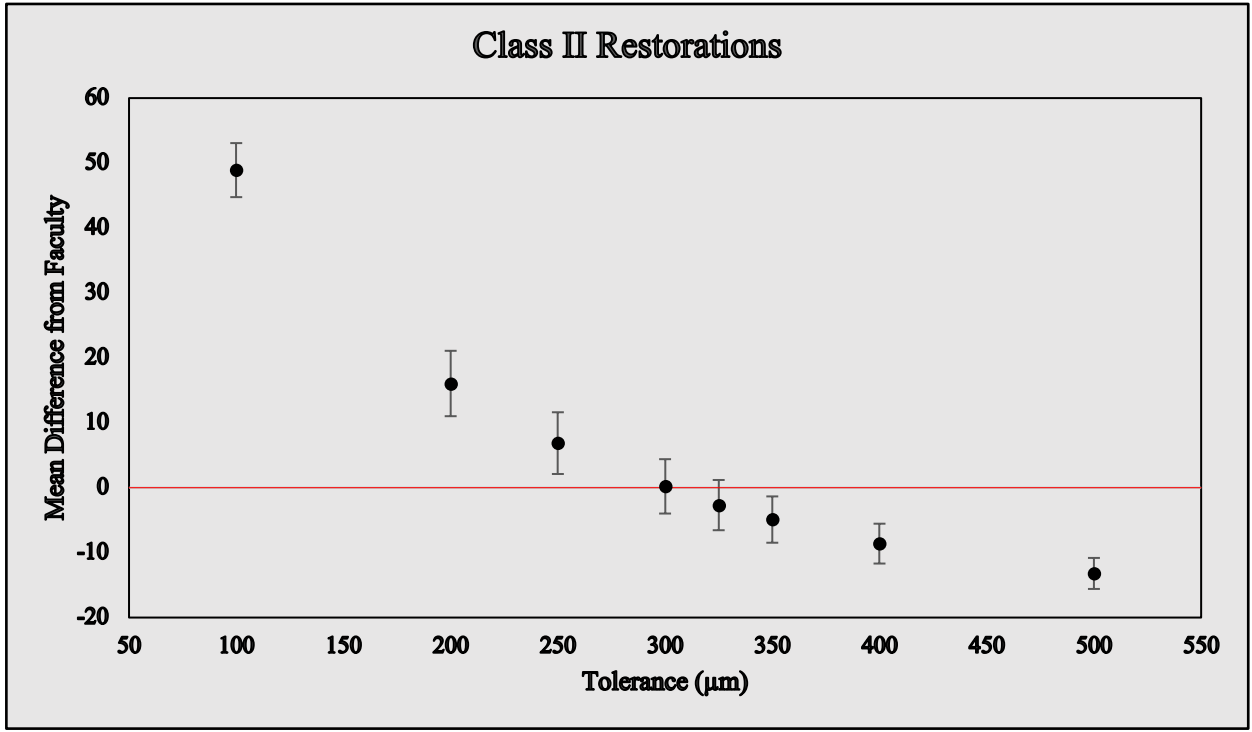

*Fig. 5*. The mean difference between faculty and machine grades for the Class II restorations at different tolerance levels.

| Tolerance Level (µm) | Class I preparation | | Class II preparation | | Class II restoration | |
| --- | --- | --- | --- | --- | --- | --- |
| | Mean | Standard Deviation | Mean | Standard Deviation | Mean | Standard Deviation |
| 100 | 32.16 | 8.03 | 25.64 | 5.18 | 31.89 | 12.11 |
| 200 | 62.82 | 9.90 | 54.20 | 8.38 | 64.75 | 16.00 |
| 250 | 71.64 | 9.67 | 63.05 | 8.80 | 73.93 | 15.09 |
| 300 | 78.82 | 8.92 | 69.95 | 8.98 | 80.59 | 13.15 |
| 350 | 84.41 | 7.88 | 75.57 | 8.59 | 85.70 | 10.89 |
| 400 | 88.80 | 6.56 | 80.05 | 7.96 | 89.41 | 9.10 |
| 500 | 95.07 | 4.52 | 85.80 | 10.42 | 94.00 | 6.41 |

*Table 1*. Statistics summary for digital grading at different tolerance levels.